\documentclass[reprint,aps,prd,nofootinbib,a4paper,10pt,twocolumn,superscriptaddress]{revtex4-2}
\usepackage{hyperref} 
\usepackage{changes}
\usepackage{graphicx,subfigure}
\usepackage[section]{placeins}
\usepackage[titletoc]{appendix}
\usepackage{xcolor}
\usepackage{dsfont}
\usepackage{bm}
\usepackage{mathtools} 
\usepackage{amsfonts}
\usepackage{enumerate}

\usepackage{dcolumn}
\usepackage{multirow}

\relax

\hypersetup{
     colorlinks   = true,
     citecolor    = cyan,
     linkcolor = pink
     }

\definecolor{MyB}{rgb}{0.1,0.1,1.0}

\definecolor{mygreen}{rgb}{0,0.5,0}

\begin{document} 

\title{Numerical evolution of Maxwell's equations for wavefronts propagating through a spherical gravitational potential}

\author{Annamalai P. Shanmugaraj}
\email{annamalaips@cp3.sdu.dk, annamalai.pshanmugaraj@gmail.com}
\affiliation{CP$^3$-Origins, University of Southern Denmark, Campusvej 55, DK-5230 Odense M, Denmark}

\author{Erik Schnetter}
\email{eschnetter@perimeterinstitute.ca}
\affiliation{Perimeter Institute for Theoretical Physics, Waterloo, ON, Canada}
\affiliation{Department of Physics and Astronomy, University of Waterloo, Waterloo, ON, Canada}
\affiliation{Center for Computation \& Technology, Louisiana State University, Baton Rouge, LA, USA}

\author{S. M. Koksbang}
\email{koksbang@cp3.sdu.dk}
\affiliation{CP$^3$-Origins, University of Southern Denmark, Campusvej 55, DK-5230 Odense M, Denmark}

\begin{abstract}
We present a scheme for numerically solving Maxwell's equations in a weakly perturbed spacetime without introducing the usual geometric optics approximation. Using this scheme, we study light propagation through a spherical perturbation of a stationary metric in the linear limit. Going beyond geometric approximation allows us to explore the wave optics effects that occur when the wavefront propagates through the gravitational potential. We find that the wavefront breaks on the gravitational potential due to the Shapiro time delay, i.e. due to the slowing down of light when propagating in an inhomogeneous spacetime. This again leads to constructive interference, resulting in a persistent frequency dependent amplification of parts of the wave amplitude. When the wave length is shorter than a certain threshold set by the induced gravitational mass, the waves are reflected on the curvature gradient.
\end{abstract}

\maketitle

\section{Introduction}\label{sec:introduction}
Our main source of information about the Universe is light that has propagated over vast distances to reach our telescopes. The light contains an abundant amount of information e.g. through redshift, spectral lines and gravitational lensing (deformation) of light ray bundles \cite{schneider1992gravitational}. In order to extract this information from the light, we need a solid theoretical understanding of its behavior. Currently, this theoretical understanding is based on the geometric optics approximation which is a low-order approximate solution to Maxwell's equations valid in the limit where the light behaves as a locally parallel plane wave with wavelength much smaller than the typical length scale over which the amplitude and polarization change as well as the typical curvature radius of spacetime along the light ray \cite{thorne2000gravitation}. It is in this approximation that light travels along null-geodesics.

The geometric optics approximation is usually extremely good. However, as telescopes are becoming increasingly precise, we are reaching a point where higher order effects such as the gravitational spin Hall effect \cite{oancea2019overview, shoom2024gravitationalfaradayspinhalleffects} may begin to be measurable. 
Several analytical endeavors to better understand beyond-geometric optics effects have been presented in the recent years. Some have been focused on general analytical schemes for describing higher order effects (see e.g. \cite{murk2024gravity, Cabral_2016, Harte_2019, Frolov_2020}), while others have focused on specific spacetimes or scenarios. The latter is for instance the case in the line of work presented in \cite{Asenjo_2017, Asenjo_2017_2, hojman2018electromagneticredshiftanisotropiccosmologies}, where the authors find analytical exact or perturbative solutions to the full Maxwell equations for specific spacetimes such as the Schwarzschild, Kerr and Gödel spacetimes, as well as anisotropic Bianchi~I spacetimes. Using these spacetimes, the authors demonstrate that light does not always propagate along null-geodesics, while the latter anisotropic spacetime is also specifically used to demonstrate that spacetime anisotropy induces birefringence and dispersion of light, affecting e.g. the observed redshift. In another paper \cite{koksbang2022effect}, the authors (including one of the current authors) develop a post-geometrical optics approximation that takes curvature terms of the general relativistic Maxwell equation into account. These terms are usually assumed sub-dominant and not considered. They may, however, be important when significant curvature gradients are present. This can for instance be the case when light travels through dark matter particles (or clumps thereof). By deriving corresponding redshift and distance relations and estimating the effects of the discrete dark matter particles on the redshift-distance relation, it was in \cite{koksbang2022effect} assessed that the effects on the distance measure are so significant that current cosmic microwave background (CMB) observations rule out dark matter masses $m\gtrsim 100$~MeV simply because the effects would otherwise be so large they should have been detected already. Putting such a low upper bound on the dark matter particle mass through cosmological observations would be very desirable, and it is therefore worth investigating the validity of the results found in \cite{koksbang2022effect}, just as it is important to further investigate the results of e.g. \cite{Asenjo_2017, Asenjo_2017_2, hojman2018electromagneticredshiftanisotropiccosmologies} to understand under what circumstances (if any) post-geometric optics effects become important for astronomical observations.
\newline\indent
Although most works going beyond the geometric optics approximation are analytical, some numerical studies do exist in the current literature. For instance, in \cite{Frolov_2012}, the authors numerically studied the trajectory of circularly polarized light propagating through a Kerr black hole using the modified geometric optics approximation introduced in \cite{Frolov_2011}. The authors observed that the trajectories of polarized photons were all null curves, coinciding with null geodesics {\em only} in the high frequency limit.
\newline\indent
To increase the feasibility of general studies into light propagation beyond the geometrical optics approximation, our goal is to be able to solve Maxwell's equations for general spacetimes. As a first step towards this, we here present the results from solving the Maxwell equations for a perturbed spacetime representing a central, spherically symmetric overdensity surrounded by vacuum. We believe that this type of numerical simulations, solving the full Maxwell equations, will be useful since they let us consider beyond-geometric optics effects at an exact level, as a complementary endeavor to the analytical studies which inevitably must introduce simplifications when the studied setups become more complicated. Beyond the possibility of setting bounds on dark matter mass, understanding beyond-geometric optics effects may prove important for correctly interpreting high precision observations of e.g. black holes (see e.g. \cite{Frolov_2012}) and for resolving ambiguous observational signatures for cosmic anisotropy (see e.g. \cite{Migkas_2020, Migkas_2021, Akarsu_2023, Blake_2002, Secrest_2021, Secrest_2022, Dam_2023, Colin_2019, Horstmann_2022, Ferreira_2021, Darling_2022}). Solving Maxwell's equations numerically would let us explore the complicated curvature effects involved without including significant assumptions or approximations. A numerical study would also enable us to validate the predictions of analytical studies.
\newline\newline
The paper is organized as follows. In section \ref{sec:theoretical_setup}, we review the geometric optics approximation before 
writing up Maxwell's equations in curved spacetime in the linear approximation of the spacetime perturbation. In section \ref{sec:numerical setup}, we describe our numerical setup and numerical methods involved in solving this set of equations including setting up initial conditions. In section \ref{sec:numerical results}, we report our findings and make comparisons with earlier works. In section \ref{sec:summary and conclusions} we conclude and discuss possible future extensions. 
\newline\indent
Throughout this paper, we adopt geometrized units with $G=1=c$, where space and time have the same units which we here adopt to be seconds.

\section{Theoretical setup}\label{sec:theoretical_setup}
Light propagation in a gravitational field is usually studied under the geometric optics approximation \cite{thorne2000gravitation} (section 22.5). In this approximation, Maxwell's equations are solved perturbatively by a locally plane-wave ansatz under the assumption that the wavelength is much smaller than the length scale over which the amplitude, polarization and radius of curvature change. The resulting equations show that light rays travel along null geodesics and are transverse. In scenarios where the length scale of curvature gradients is comparable to the wavelength of light involved, the geometric optics approximation is expected to fail. As discussed in \cite{koksbang2022effect}, this situation might for instance be relevant in the real universe because non-fussy dark matter particles in principle can lead to steep local curvature gradients. We use the considerations of \cite{koksbang2022effect} as the motivation for specifying the spacetime we consider here. To further study the possible effects of local curvature gradients on light propagation we will thus consider light propagation in a spacetime describing a central overdensity surrounded by vacuum. Below, we introduce the metric and corresponding Maxwell equations.
\newline\newline
We will consider the spacetime in the first order perturbed limit where the line element can be written as
\begin{equation}\label{eq1}
    ds^2 = -(1-2\alpha) dt^2 + (1+2\alpha)(dx^2 + dy^2 + dz^2),
\end{equation}
where $\alpha$ is the scalar perturbation. We require the perturbation to correspond to a central overdensity given by
\begin{equation}
    \rho(r) = \frac{M}{ \lambda^3 (2\pi)^{3/2}  } \cdot e^{ -r^2/(2 \lambda^2)},
\end{equation}
where $\lambda$ is a constant that determines the width of the overdensity and $M$ is the total gravitational mass in the region. Inspired by \cite{koksbang2022effect} we will consider the overdensity as modeling the classical mass distribution of a dark matter particle. We therefore set $\lambda = \lambda_\mathrm{pf} \frac{2 \pi}{M} \ll 1$, where $\lambda_\mathrm{pf}$ is a prefactor that will let us set the particle size as a scaling of the Compton wavelength, $\frac{2 \pi}{M}$.
\newline\newline
We obtain the scalar metric perturbation $\alpha$ by analytically solving Poisson's equation $-\nabla^2 (\alpha) = 4 \pi \rho(r)$ which yields
\begin{equation}
    \alpha(r) = \frac{M}{r} \cdot \mathrm{erf}\left(\frac{r}{\sqrt{2} \lambda} \right),
\end{equation}
where $\mathrm{erf}$ is the error function. For $r\gg\lambda$, the error function approaches unity, and the scalar perturbation $\alpha(r)$ approaches the point-mass potential
\begin{equation}
    \alpha(r) \simeq \frac{M}{r} .
\end{equation}
We require $\alpha$ and its derivatives to be numerically much smaller than 1 so that we can work in the linear limit. Specifically, we will require the maximum amplitude of $\alpha$ to be $\alpha_{max} = 0.1$ and require the maximum of the relevant derivatives of $\alpha$ to be of this order as well. This choice determines the mass of the dark matter particle according to
\begin{equation}
\begin{split}
    M = \alpha_\mathrm{max} \lambda \sqrt{\frac{\pi}{2}} &= 2 \pi \alpha_\mathrm{max} \lambda_\mathrm{pf} \sqrt{\frac{\pi}{2}} , \\
    M & = \sqrt[4]{2 \pi^3} ( \lambda_\mathrm{pf} \alpha_\mathrm{max})^{1/2}.
\end{split}
\end{equation}
Note that at $r=0$, $\alpha$ takes an indeterminate form. To evaluate the functions at that point, we therefore use L'Hôpital's rule which yields
\begin{equation}
        \alpha |_{r=0} = \frac{M}{\lambda} \sqrt{\frac{2}{\pi}}.
\end{equation}
From this, we can compute the first and second derivatives of $\alpha$ numerically.
\newline\indent
To obtain the correct Maxwell equations linearized in $\alpha$ we need the full, nonlinear Christoffel symbols which are
\begin{equation*} 
  \Gamma^{\lambda}\,_{\rho \sigma} =
    \begin{cases}
    \Gamma{}_{t}{}_{i}{}^{t}= \partial_{i}{\alpha} ( -1 + 2\alpha)^{-1}\\[-.5ex]
    
    \Gamma{}_{y}{}_{y}{}^{y} = \Gamma{}_{x}{}_{y}{}^{x} = \Gamma{}_{z}{}_{y}{}^{z} =  \partial_{y}{\alpha} ( 1 + 2\alpha)^{-1}\\[-.5ex]

    \Gamma{}_{y}{}_{x}{}^{y} = \Gamma{}_{x}{}_{x}{}^{x} = \Gamma{}_{z}{}_{x}{}^{z}= \partial_{x}{\alpha} ( 1 + 2\alpha)^{-1}\\[-.5ex]
    
    \Gamma{}_{z}{}_{y}{}^{y} =\Gamma{}_{z}{}_{x}{}^{x} =  \partial_{z}{\alpha} ( 1 + 2\alpha)^{-1}\\[-.5ex]

    \Gamma{}_{z}{}_{z}{}^{x} = \Gamma{}_{y}{}_{y}{}^{x}= \Gamma{}_{t}{}_{t}{}^{x} = -\partial_{x}{\alpha} ( 1 + 2\alpha)^{-1}\\[-.5ex]
    \Gamma{}_{z}{}_{z}{}^{y}= \Gamma{}_{x}{}_{x}{}^{y} = \Gamma{}_{t}{}_{t}{}^{y} = -\partial_{y}{\alpha} ( 1 + 2\alpha)^{-1}\\[-.5ex]
    \Gamma{}_{y}{}_{y}{}^{z}= \Gamma{}_{x}{}_{x}{}^{z} = \Gamma{}_{t}{}_{t}{}^{z} = -\partial_{z}{\alpha} ( 1 + 2\alpha)^{-1}\\[-.5ex]

    \end{cases}       .
\end{equation*}
With these at hand, we can now set up Maxwell's equations for the spacetime in and around the region mimicking a classical dark matter particle.

\subsubsection*{Maxwell's equations in curved spacetime}
Maxwell's equations can be derived by varying the action
\begin{equation}
    S = - \frac{1}{4}\int d^4x \sqrt{-g} F_{\rho \sigma} F^{\rho \sigma},
\end{equation}
where $F_{\rho \sigma}$ is the electromagnetic field tensor which can be written in terms of the four-vector potential, $A_{\sigma}$, according to $F_{\rho \sigma} = \partial_{\rho} A_\sigma - \partial_\sigma A_\rho$. Assuming that the energy of the electromagnetic field is small compared to the energy of the dark matter particle, we can ignore its energy contribution. This results in the source-free tensor equation
\begin{equation}
    -\nabla_\rho \nabla^\rho A^\sigma + R^\sigma _\rho A^\rho + \nabla^{\sigma} (\nabla _\rho A^\rho) = 0.
\label{maxwell eqn}
\end{equation}
We will start by expanding the first term using the metric introduced above. Writing  out the terms using partial derivatives and Christoffel symbols, we get
\begin{align}
\begin{split}
      -\nabla_\rho \nabla^\rho A^\sigma &=\\ &- \bigl(\partial_{\rho}{g^{\rho \gamma} \partial_{\gamma}{ A^{\sigma} }} 
            + g^{\mu \rho} A^{\gamma} \partial_{\rho}{ \Gamma^{\sigma}_{\gamma \mu}}
		+ g^{\mu \rho} \Gamma^{\sigma}_{\gamma \mu} \partial_{\rho}{ A^{\gamma}}\\
		& + \Gamma^{\rho}_{\delta \rho} g^{\mu \delta} \partial_{\mu}{ A^{\sigma}}
		+ \Gamma^{\sigma}_{\delta \rho} g^{\mu \rho} \partial_{\mu}{A^{\delta}}\bigr),
\end{split}
\end{align}
where we have ignored non-linear terms in $\alpha$ and its derivative which e.g. lets us neglect products of the Christoffel symbols.
\newline\indent
The Ricci tensor can be computed from the Christoffel symbols using the relation
\begin{equation}
    R_{\mu \nu} = \partial_\rho \Gamma^\rho_{\mu \nu} - \partial_\nu \Gamma^\rho_{\mu \rho} + \Gamma^\rho_{\mu \nu} \Gamma^\sigma_{\rho \sigma} - \Gamma^\rho_{\mu \sigma} \Gamma^\sigma_{\nu \rho}.
\end{equation}
Again ignoring the non-linear terms in $\alpha$ and its derivatives, the second term in the Maxwell equation can be expanded as
\begin{equation}
    R^\nu _{\mu} A^\mu = g^{\nu \sigma} ( \partial_\rho \Gamma^\rho_{\mu \sigma} - \partial_\sigma \Gamma^\rho_{\mu \rho} ) A^\mu.
\end{equation}
Inserting $\nu=t,i$, this gives the expressions
\begin{align}
\begin{split}
    R^t _{\mu} A^\mu &= \alpha_{,jj}A^t\\
    R^i _{\mu} A^\mu &= -\alpha_{,jj}A^i.
\end{split}
\end{align}
\newline
For the last term in Maxwell's equation, $\nabla^{\sigma} (\nabla _\rho A^\rho)$, we will introduce the Lorenz gauge condition, which implies that $\nabla _\rho A^\rho = 0$. Hence, the last term is set to zero. The Lorenz gauge will function as the constraint in our system and we will monitor the constraint violation throughout the numerical evolution.
\newline\indent
Combining the above, the full Maxwell's equations in curved spacetime for the linearly perturbed Minkowski metric in the Lorentz gauge are found to be
\begin{widetext}

\begin{align}
\begin{split}
     (1 + 2 \alpha) \partial_{tt} \phi -(1 - 2 \alpha) [ \partial_{xx} \phi + \partial_{yy} \phi + \partial_{zz} \phi ] \\ + 2(\alpha_{,xx} + \alpha_{,yy} + \alpha_{,zz}) \phi  - 2( \alpha_{,x} \partial_t A^x + \alpha_{,y} \partial_t A^y + \alpha_{,z} \partial_t A^z ) \\ + 2 (\alpha_{,x} \partial_x \phi + \alpha_{,y} \partial_y \phi + \alpha_{,z} \partial_z \phi) &= 0,
     \label{phi}
\end{split}
\end{align}

\begin{align}
\begin{split}
    (1 + 2 \alpha) \partial_{tt} A^x -(1 - 2 \alpha) [ \partial_{xx} A^x + \partial_{yy} A^x + \partial_{zz} A^x ] \\ - 2(\alpha_{,xx} + \alpha_{,yy} + \alpha_{,zz}) A^x - 2( \alpha_{,z} \partial_x A^z + \alpha_{,y} \partial_x A^y) \\ + 2 \alpha_{,x} ( -\partial_t \phi + \partial_y A^y + \partial_z A^z ) - 2 (\alpha_{,x} \partial_x A^x + \alpha_{,y} \partial_y A^x + \alpha_{,z} \partial_z A^x ) &= 0,
    \label{Ax}
\end{split}
\end{align}

\begin{align}
\begin{split}
    (1 + 2 \alpha) \partial_{tt} A^y -(1 - 2 \alpha) [ \partial_{xx} A^y + \partial_{yy} A^y + \partial_{zz} A^y ] \\- 2(\alpha_{,xx} + \alpha_{,yy} + \alpha_{,zz}) A^y - 2( \alpha_{,x} \partial_y A^x + \alpha_{,z} \partial_y A^z) \\ + 2 \alpha_{,y} ( -\partial_t \phi + \partial_x A^x + \partial_z A^z ) - 2 (\alpha_{,x} \partial_x A^y + \alpha_{,y} \partial_y A^y + \alpha_{,z} \partial_z A^y )  &= 0,
    \label{Ay}
\end{split}
\end{align}

\begin{align}
\begin{split}
    (1 + 2 \alpha) \partial_{tt} A^z -(1 - 2 \alpha) [ \partial_{xx} A^z + \partial_{yy} A^z + \partial_{zz} A^z ] \\ - 2(\alpha_{,xx} + \alpha_{,yy} + \alpha_{,zz}) A^z - 2( \alpha_{,x} \partial_z A^x + \alpha_{,y} \partial_z A^y) \\ + 2 \alpha_{,z} ( -\partial_t \phi + \partial_x A^x + \partial_y A^y ) - 2 (\alpha_{,x} \partial_x A^z + \alpha_{,y} \partial_y A^z + \alpha_{,z} \partial_z A^z ) &= 0,
    \label{Az}
\end{split}
\end{align}

\end{widetext}
where we have used the notation $\partial_{i} \alpha = \alpha_{,i} $ and we have substituted $\phi$ in place of $A^t$ to denote the scalar electromagnetic potential in accordance with standard notation. We thus denote the four variables of the four coupled linear partial differential equations $\{\phi, A^x,A^y,A^z\}$.
\newline\newline
We also define a set of wave equations that involve only non-covariant terms $g^{\gamma \mu} \partial_{\mu} \partial_{\gamma} A^\nu$, i.e.,
\begin{equation}
    \left[ (1 + 2 \alpha)\partial_{tt} - (1 - 2 \alpha)( \partial_{xx} + \partial_{yy} + \partial_{zz}) \right] A^\mu = 0
\end{equation}
We will evolve these ``simplified'' wave equations and compare their solutions with those of the full wave equations \eqref{phi}, \eqref{Ax},\eqref{Ay} and \eqref{Az}. This will allow us to probe the contributions from the covariant (curvature) terms.
\newline\newline
The goal is now to evolve these equations numerically for a setup with initially locally plane waves propagating towards and through a dark matter particle/spherical mass distribution.

\section{Numerical setup} \label{sec:numerical setup}
In this section, we describe our numerical setup for two simulations corresponding to the evolution of the partial differential equations derived above. The two simulations differ only by the value of the wave vector $|\vec{k}|$.
\newline\indent
In the first subsection below, we describe the evolution of Maxwell's equations in 1+3 dimensions. We then move on to describe the initial conditions for the simulations. In the final subsection, we describe parameters that influence simulation feasibility and use this to specify the parameters of the two simulations we run.

\subsection{Evolution in 1+3D}
The Maxwell equations in curved spacetime are a set of four coupled second order linear hyperbolic partial differential equations (PDEs). By substitution, we convert the second order time derivatives into first order derivatives, yielding eight coupled first order PDEs. These equations can be evolved in time using a fourth order Runge Kutta time integrator. To compute the gradients and divergences appearing in the eight PDEs, we use the Method of Lines \cite{scheisser1991numerical} to discretize in all but one dimension. This leads to a system of ordinary differential equations (ODEs) where we can use a fourth order central finite difference stencil to compute the derivatives. We use Einstein Toolkit's CarpetX framework \cite{schnetter2010carpetx,Löffler_2012} to evolve this set of eight coupled space- and time-discretized ODEs.
\newline\newline
We are only interested in the simulation domain interior and do not want boundary effects to influence the interior of our simulation. The simplest choice is thus to use either periodic or reflecting boundary conditions. Here, we use periodic boundary conditions in the $x$ and $y$ coordinates as the waves are symmetric and move perpendicular to this direction. We use reflecting boundaries in the $z$ coordinate 
and stop the simulation when the wave reaches the end of the simulation box to prevent reflected waves from affecting our results.

To assess the accuracy of our results, we track the constraint violation in our system at all times, i.e. we track the fulfillment of the Lorenz gauge condition, 
\begin{equation}
    \begin{split}
        \nabla_\alpha A^\alpha & = \partial_\alpha A^\alpha + \Gamma^\alpha _{\alpha \beta} A^\beta \\
        & = ( \partial_t A^t + \partial_x A^x + \partial_y A^y + \partial_z A^z) \\ &+ 2( \alpha_{,x} A^x + \alpha_{,y} A^y + \alpha_{,z} A^z).
    \end{split}
    \label{lorenz}
\end{equation}

\subsection{Initial conditions}
\begin{figure}
\includegraphics[width=8cm]{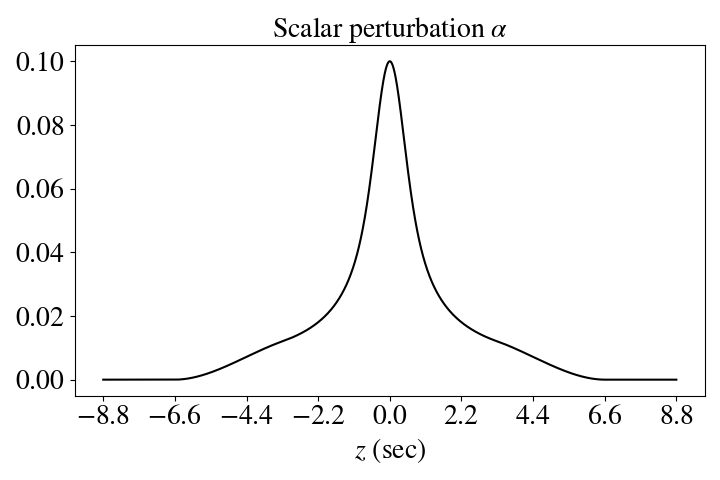}
\includegraphics[width=8cm]{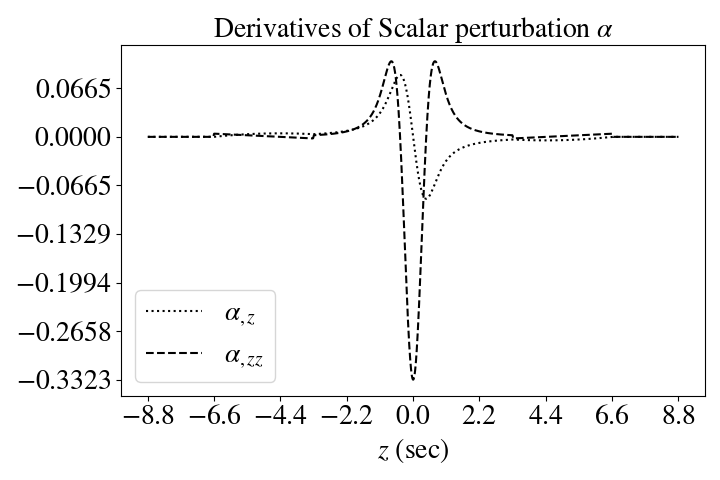}
\caption{Scalar metric perturbation $\alpha$ and its derivatives with $\lambda_\textrm{pf} = 0.002$. The perturbation includes a spline approximation. This is for the simulation with $|\Vec{k}| = 1.5 M_\textrm{gi}$, where a spline was introduced to force $\alpha$ to zero at a finite distance. Spatial distances are in units of seconds.
}\label{fig:alpha}
\end{figure}

We consider a linearly polarized monochromatic plane wave propagating along the z-axis as our initial condition. In vacuum, the vector potential $A^i$ takes the form
\begin{equation}\label{eq:planewave}
    \begin{split}
        A^x & = \mathcal{A} \cos[ 2 \pi (k_zz-\omega t)] \\
        A^y & = 0 \\ 
        A^z & = 0,
    \end{split}
\end{equation}
where $\mathcal{A}$ is the amplitude of the wave, $\vec{k}$ is the spatial wave vector and $\omega = \sqrt{k_x^2+k_y^2+k_z^2} = |\vec{k}|$ is the energy of the light wave.
\newline\indent
It is well-known that the above plane waves are not in general a solution of Maxwell's equations in a gravitational field (\cite{schneider1992gravitational}, page 93). We must therefore initialize our waves to have non-vanishing amplitude only in the region where the curvature is very weak, i.e. far from the central overdensity/dark matter particle. This can for instance be achieved by considering a plane wave packet propagating towards and through the dark matter particle from infinity. However, it is not practically feasible to initialize the plane wave sufficiently far from the origin for $\alpha$ to be negligible; this would require a very large simulation domain with high grid resolution and long time evolution. We therefore modify our scalar metric perturbation $\alpha$ by a spline that forces $\alpha$ to vanish identically at a finite radius. Specifically, we insert a spline approximation at the distance from the origin where $\alpha$ reaches $12\%$ of its maximum value and force $\alpha$ to zero where $\alpha$ would otherwise have reached $6\%$ of its maximum value. Mathematically this can written as
\begin{equation}
    \alpha(r)= 
        \begin{cases}
    \frac{M}{r} \cdot \mathrm{erf}\left(\frac{r}{\sqrt{2} \lambda} \right), & \text{if } r\leq r_1\\
    \text{Spline approximation},              & \text{if } r_1 \leq r \leq r_2 \\
    0, & \text{if } r \geq r_2,
\end{cases}
\end{equation}
where $r = \sqrt{x^2 + y^2 +z^2}$ and $r_1$, $r_2$ correspond to coordinate points where $\alpha$ reaches $12\%$ of $\alpha_{max}$ and $6\%$ of $\alpha_\mathrm{max}$, respectively. The resulting $\alpha$ and its derivatives are shown in figure \ref{fig:alpha}.
\newline\indent

By forcing $\alpha$ and its derivatives to vanish at finite $r$, we are effectively rendering the central overdensity gravitationally ``invisible''. This is only possibly by surrounding the overdensity by an underdensity which here corresponds to a negative density contribution. This is clearly not physically meaningful. However, the introduction of the spline should be considered a purely {\em numerical} approximation introduced to mimic initializing the light very far from the central overdensity\footnote{We thank Syksy Rasanen for this suggestion.}. There will be no artifacts of this approximation since light does not ``care'' that it propagates through negative density, i.e. the equations for light propagation do not exhibit any striking behavior due to negative density. Nonetheless, in order to verify that the spline approximation does not introduce unforeseen artifacts, we have compared our results with simulations without the spline and hence without negative density. These results show overall agreement with the results obtained with the spline, i.e. the results we present below are the same as those we obtained without the spline approximation. The main difference between the two simulations is that when the spline is not introduced, the Lorenz gauge constraint is violated initially because we use initial conditions corresponding to $\alpha = 0$. Although we could modify the initial conditions, this is non-trivial since it requires careful consideration not to introduce spurious gauge modes. This is the main reason we prefer introducing the spline, again emphasizing that it is meant as a numerical approximation for mimicking that the light propagates in from infinity.
\newline\newline
With the above in place, we initialize the wave at a distance slightly greater than $r_2$ where $\alpha = 0$. More precisely, we initialize the wave as a wave packet by introducing the wave amplitude
\begin{equation}
    \mathcal{A} = \left(  \frac{1}{1 + e^{-2q(z-l_1)}} + \frac{1}{1 + e^{-2q(l_2 - z)}} -1  \right),
\end{equation}
where $l_1$ and $l_2$ are the distances inside which wave fronts are confined at initial time. The parameter $q$ determines how smoothly the wave fronts reach their peak amplitude. For very small $q$, the wave packet would look like a Gaussian pulse. For bigger values of $q$, we would instead have a square wave function.
\newline\indent
Since the wave amplitude at initial time is only non-negligible in regions with $\alpha = 0$, we may, as discussed above, use the plane wave of equation (\ref{eq:planewave}) together with $\phi = 0$ as our initial conditions without breaking the Lorenz gauge condition.

\subsection{Final specification of numerical simulations}
We have set up our simulations such that we have two parameters that can be varied: The ``size'' of the dark matter particle ($\lambda$) and the wave vector $|\vec{k}|$. When choosing model parameters, we will specify two different simulations based on the result from \cite{koksbang2022effect} where it was found that when light propagates through curved spacetime, the light acquires a gravitationally induced mass given by the relation
\begin{equation}
    M_\textrm{gi} \equiv \sqrt{\frac{\rho - p}{2M_{pl}^2}} = \sqrt{4 \pi (\rho - p)},
\end{equation}
where $\rho$ and $p$ are the density and pressure of the dark matter particle. In the analytical investigation \cite{koksbang2022effect}, it was found that for light to propagate from vacuum into the matter region, $|\vec{k}| > M_\textrm{gi}$ must be fulfilled. Otherwise, the light cannot enter the region.
\newline\indent
For the model considered here, we can use Einstein's equation to write 
\begin{equation}
    p = \frac{\rho - R}{3}, \hspace{3mm} R = -2 \sum_i \alpha_{,ii}.
\end{equation}
When using this equation below, we will calculate $M_\textrm{gi}$ at the center of the dark matter particle where $\alpha$ is at its maximum. We will consider two simulations. In one, we will choose parameters such that $|\vec{k}| > M_\textrm{gi}$ while we choose $|\vec{k}| < M_\textrm{gi}$ for the other simulation. 
\newline\newline
For the simulations to be numerically feasible, we must require the scales of all features of the simulation to be of the same order of magnitude. This is because a quantity with features (like peaks or troughs) can be resolved only for a certain grid resolution and above. If we introduced several scales into the simulations, we would need to use simulations large enough to be sensible for the largest scales of our setup while still having a resolution high enough for the smallest scales of the setup. Specifically, we therefore require the width of $\alpha$ to be of the same order as the wavelength of the propagating wave. When choosing parameters we also take into consideration that we want the distribution of $\alpha$ to be narrow so that the Ricci scalar is as large as possible in order to observe the most pronounced effects of steep curvature gradients. At the same time, we must make sure that first and second derivatives of $\alpha$ are smaller than 1 so that the linear approximation still holds. As mentioned earlier, we will choose $\alpha_{\rm max} = 0.1$. Therefore, we set $\lambda_\textrm{pf} = 0.002$ which fixes the curvature to the order $1$. By specifying $\lambda_\textrm{pf}$, we automatically also fix $M_\textrm{gi}$. If we use values of the curvature greater than $1$ (or $\lambda_\textrm{pf}<10^{-3}$), then we will only be able to perform simulations with $|\vec{k}| < M_\textrm{gi}$ because the wavelength would otherwise be too small compared to the curvature and thus require very large domain resolution. We empirically find that the width of $\alpha$ is comparable to the light's wavelength when we choose $[0.1 M_\textrm{gi} < |\vec{k}| < 1.5 M_\textrm{gi}]$. Based on this, we make two simulations with parameters specified in table \ref{table:sim_parameters}. One simulation has $|\vec{k}| = 0.1M_\textrm{gi}$ while the other has $|\vec{k}|  = 1.5M_\textrm{gi}$.

\begin{table}
    \centering
    \begin{tabular}{|c|c|c|c|c| }
    \hline
         \multirow{2}{*}{Parameter} & \multicolumn{2}{|c|}{Simulations} \\

          & $|\Vec{k}|>M_\textrm{gi}$ & $|\Vec{k}|<M_\textrm{gi}$  \\
         \hline
         3D grid size & $128^2_{xy} \times 2048_z$ & $128^2_{xy} \times 2048_z$   \\
         CFL factor & $0.5$ & $0.5$  \\
         $\lambda_\textrm{pf}$ & $2 \times 10^{-3}$& $2 \times 10^{-3}$ \\
         $|\Vec{k}|$ & $ 1.5M_\textrm{gi}$ & $ 0.1M_\textrm{gi}$ \\
         $\lambda_{\text{photon}}$ & $\sim 1.4 [s]$ & $\sim 21 [s]$ \\
         \hline
    \end{tabular}
    \caption{Parameters specifying the two simulations.}\label{table:sim_parameters}
\end{table}

\section{Numerical results} \label{sec:numerical results}
In this section, we present the results of our simulations and compare them to earlier work. We discuss the results of the two simulations separately, beginning with the simulation with $|\vec{k}| > M_\textrm{gi}$. Convergence tests for the simulations are shown in appendix \ref{app:convergence}.

\begin{figure*}
\includegraphics[width=13cm]{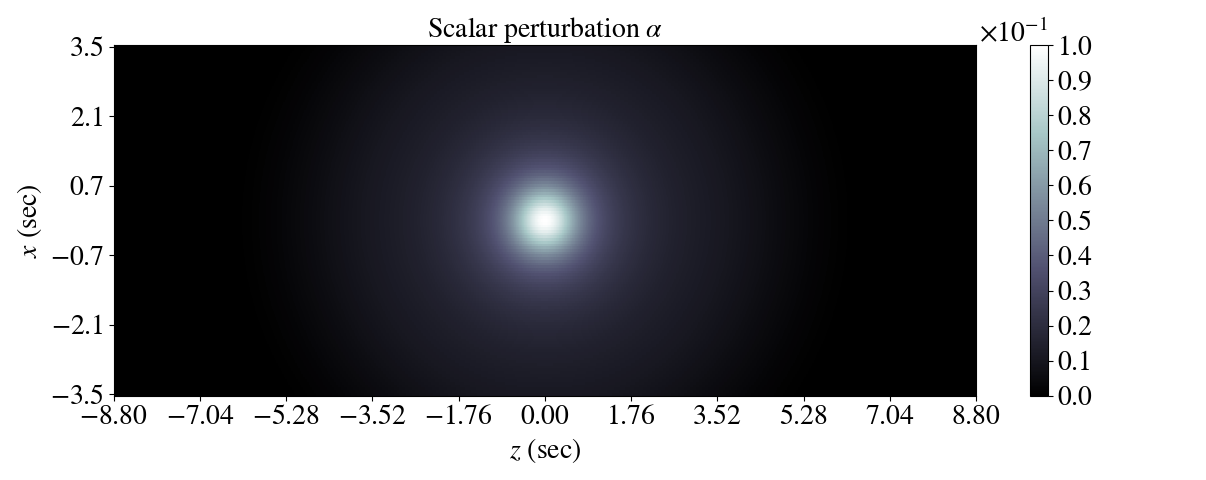}
\includegraphics[width=13cm]{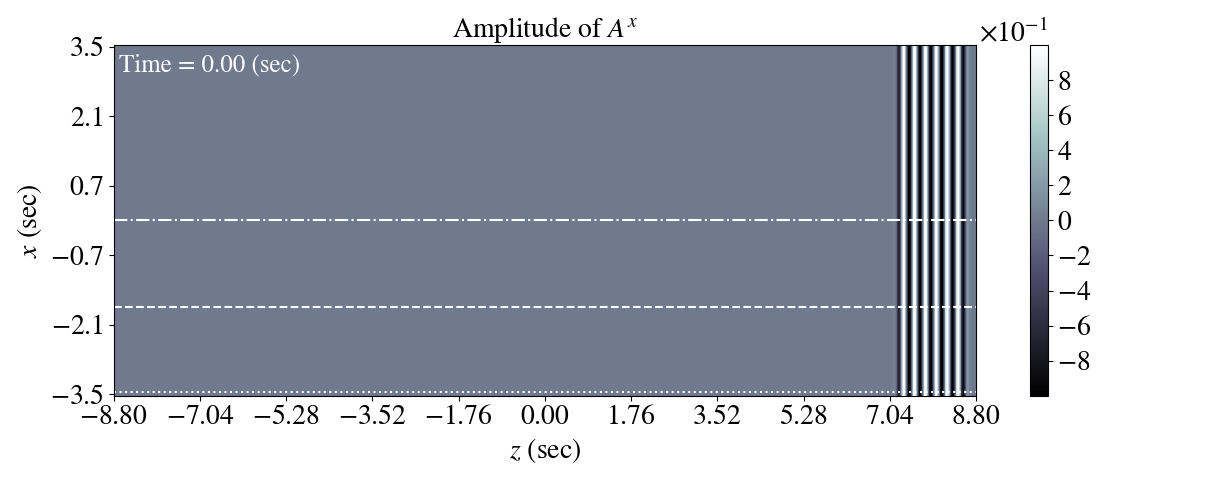}
\includegraphics[width=13cm]{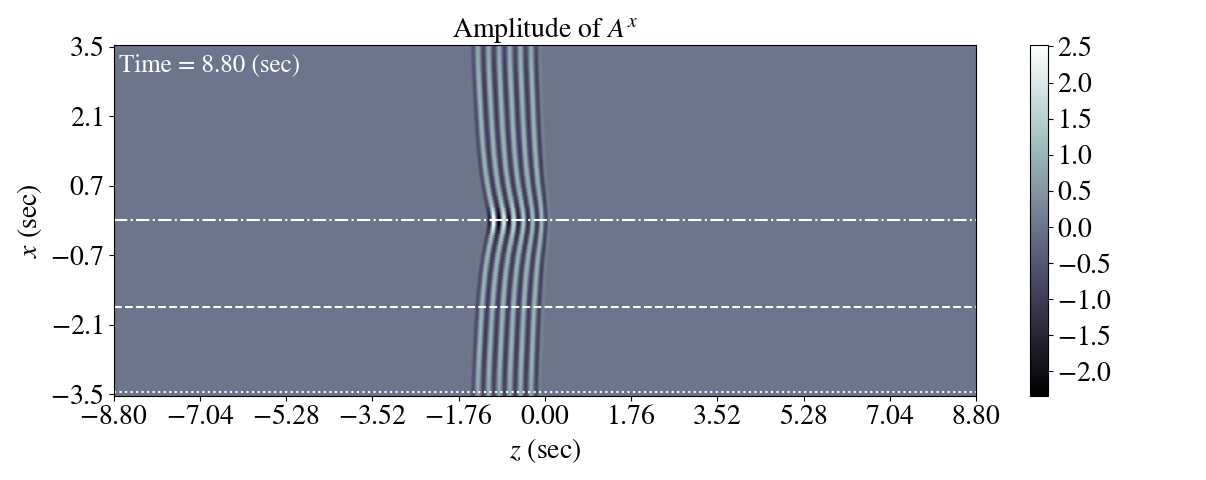}
\includegraphics[width=13cm]{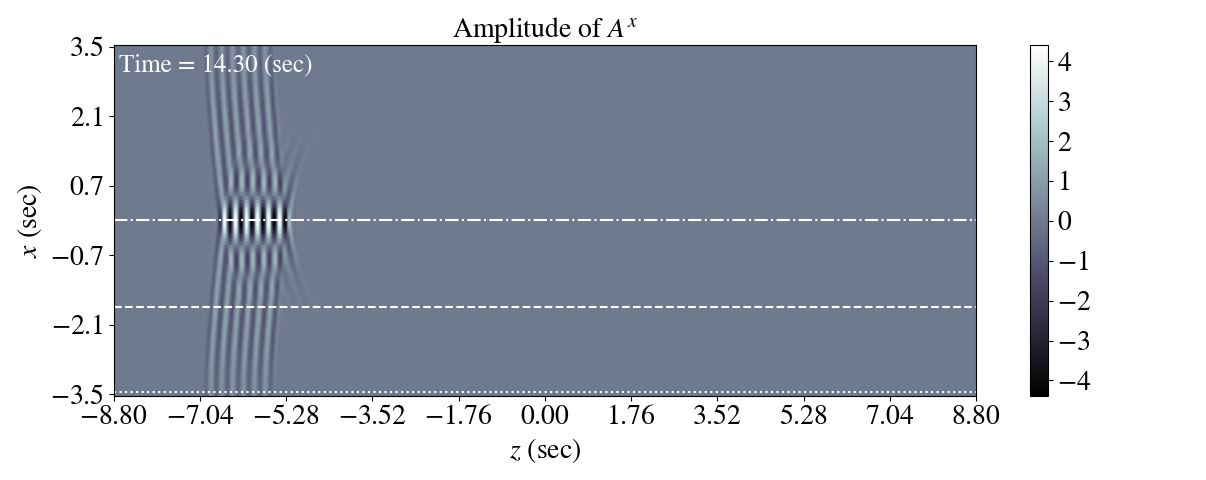}
\caption{The top figure shows the scalar perturbation $\alpha$ in the $xz-$ plane. The bottom three plots show the evolution of $A^x$ for the simulation with $|\Vec{k}| = 1.5M_\textrm{gi}$. The plots include three horizontal lines parallel to the z-axis and at different distances in the x-axis, indicating where we track the maximum amplitude of wave-fronts as shown in figure \ref{amp,kgM}.}
\label{evol_kgM}
\end{figure*}

\subsection{Simulation with $|\vec{k}| = 1.5 M_\textrm{gi}$}
Figure \ref{evol_kgM} shows the amplitude of $A^x$ at three time stamps of the simulation: At initial time, when the entire wavefront has passed through the origin, and when the wavefront is again ``far'' from the dark matter particle. It is clearly seen that the wave is affected when it propagates through the dark matter particle, resulting in interference of the wave fronts. This interference leads to a dramatic increase in the amplitude of the central part of the wave, where the amplitude increases by around a factor of 4 compared to the initial amplitude. This increase in amplitude persists even after the wave has traversed the inhomogeneous region, indicating that the effect will accumulate if light propagates through a region with multiple overdensities (such as multiple discrete dark matter particles). Determining the exact form of the interference pattern from propagation through multiple overdensities/particles would of course require considering a simulation with multiple structures.
\newline\indent
Each snapshot in figure \ref{evol_kgM} also shows three lines parallel to the $z$-axis. Figure \ref{amp,kgM} shows the amplitude of the wavefront along these three lines. The figure highlights the significant change in amplitude in the part of the wave that propagates through the central part of the overdensity.
\newline\newline
The interference pattern seen in figure \ref{evol_kgM} stems from the curvature affecting the speed of the light wave. To show that the speed of light is altered by the inhomogeneity of spacetime we can compute the refractive index as \cite{koksbang2022effect} 
\begin{equation}
    n = \left(1-\frac{M_\textrm{gi}^2}{E^2}\right)^{-1/2},
\end{equation}
where we have $E = |\vec{k}| = 1.5M_\textrm{gi}$. We thus get $n = \left(1-\frac{1^2}{1.5^2}\right)^{-1/2} \simeq 1.34 $, giving a velocity $v = c/n\approx 0.75c$ at the center of the particle/overdensity. (For comparison, one may note that the refractive index of water is $\sim 1.33$.) From this we expect that the part of the wave propagating through the most central part of the overdensity is retarded the most which is exactly what we see in figure \ref{evol_kgM}. This slowing down of light in gravitational potential wells is well-known in the literature as the Shapiro time delay \cite{PhysRevLett.13.789}.
\newline\newline\noindent

\begin{figure}
    \centering
    \includegraphics[width=1\linewidth]{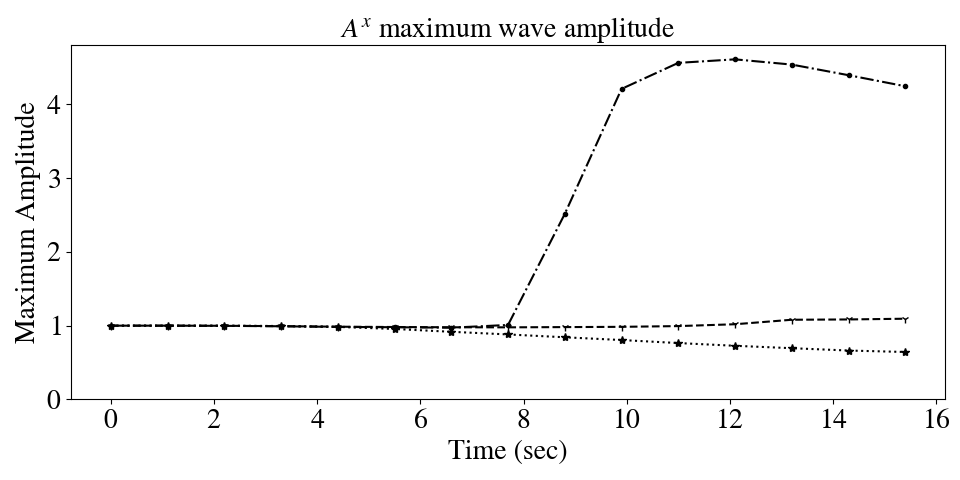}
    \caption{The maximum amplitude of the waves at different distances from the x-axis as it propagates through the dark matter particle for the simulation with $|\Vec{k}| = 1.5 M_\textrm{gi}$. The different line types correspond to the horizontal lines in figure \ref{evol_kgM} where we track the maximum amplitude of wave-fronts. The amplitude along the dot-dashed line at $x=0$ has increased while the dotted line (corresponding to the negative x-axis boundary) has decreased slightly.}
    \label{amp,kgM}
\end{figure}
\begin{figure*}
\includegraphics[width=13cm]{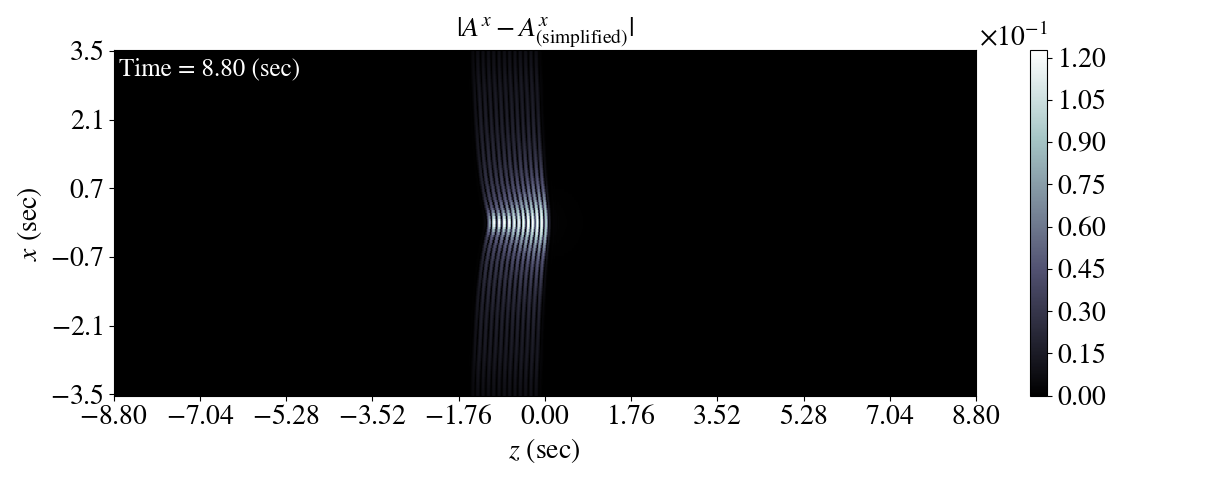}
\includegraphics[width=13cm]{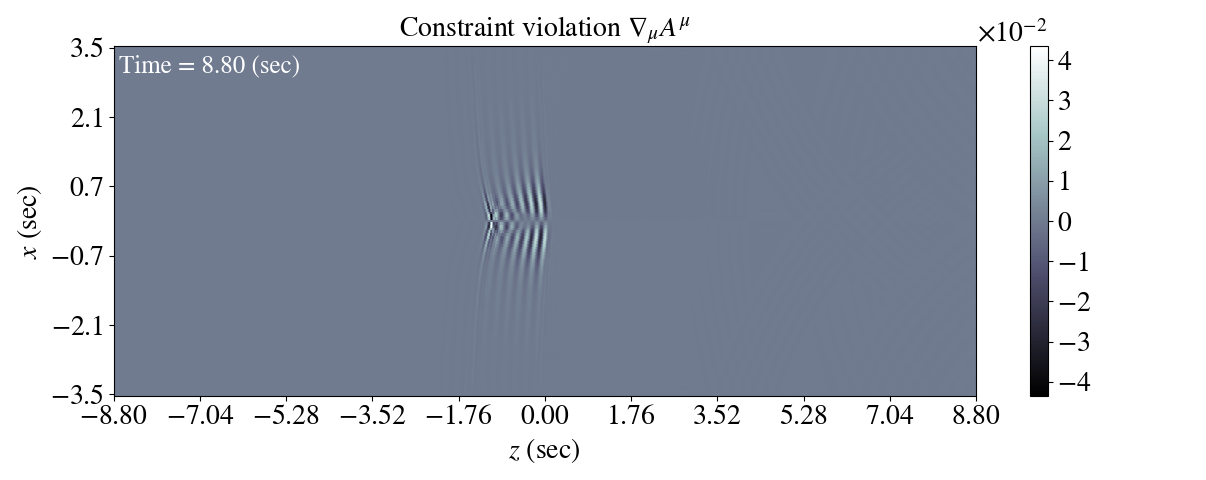}
\includegraphics[width=13cm]{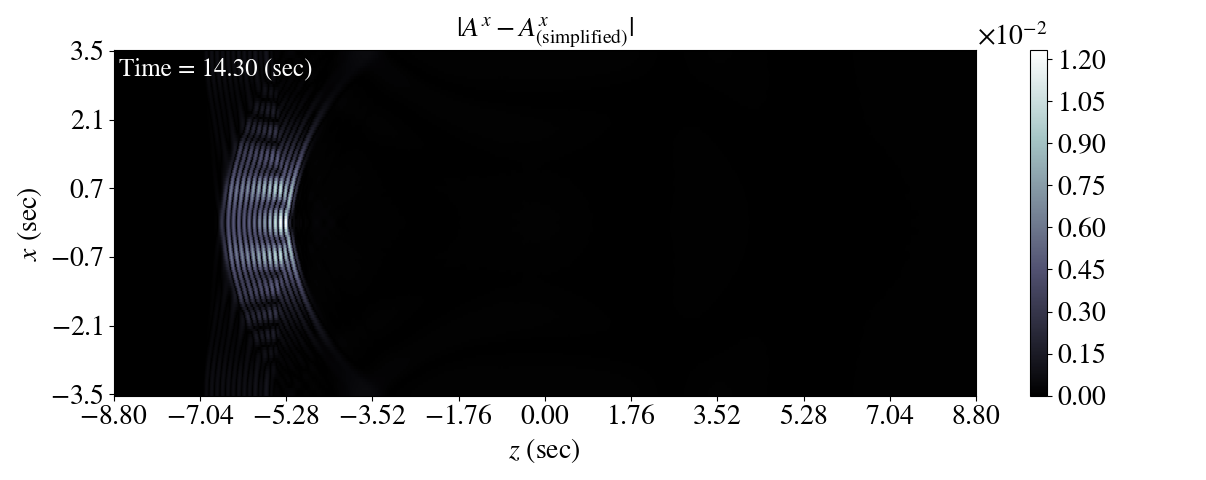}
\includegraphics[width=13cm]{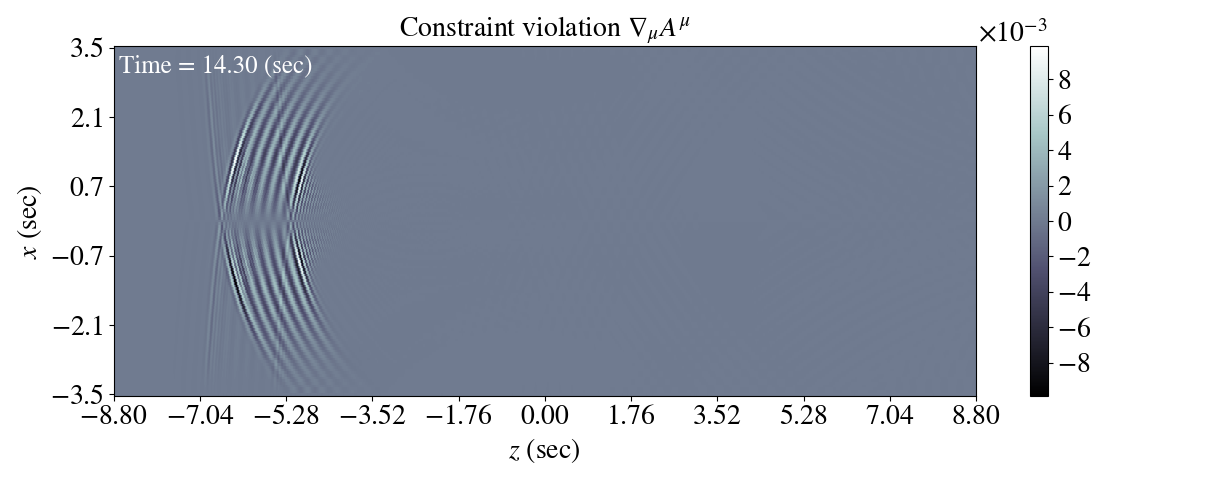}
\caption{The absolute difference between $A^x$ (eq. \eqref{Ax}) and simplified wave equation $A^x$, and its corresponding constraint violation at two time instances for the simulation with $|\vec{k}| = 1.5M_\textrm{gi}$. The center of the dark matter particle is at the origin. Constraint violation plots are for simulations with full Maxwell's equation.}
\label{nc_kgM}
\end{figure*}

When solving the simplified Maxwell equations we obtain a similar interference pattern, indicating that the interference pattern is simply an effect of standard wave optics and not specific to including the Ricci-term in Maxwell's equations. To see if there is any significant difference between the two solutions we compute their absolute difference. The results are shown in figure \ref{nc_kgM}. The difference between the two sets of results is shown at two time instances. The figure also shows plots of the constraint violation at the same time instances. At the time instance where the entire wavefront has just passed through the Origin, we see that the covariant terms have modified the waveform at the order of  $10^{-1}$ compared to the simplified wave equation solution. Since we are working in the linear regime with $\alpha = 0.1$ and the derivatives of $\alpha$ of the same order, we would naively expect our results to be valid up to the order $10^{-2}$. We therefore expect a deviation between the full and simplified Maxwell equation results at 10\% to be a genuine result and not an artifact of our approximations. We also note that the constraint violation (shown in figure \ref{nc_kgM}) is of order $10^{-2}$. We cannot expect a smaller constraint violation than this since our equations are only correct up to linear order in $\alpha\sim 10^{-1}$ and thus have errors expected to be of order $\alpha^2 \sim 10^{-2}$.
At the later time stamp shown in figure \ref{nc_kgM}, the wave fronts have passed through the dark matter. At this point, the difference between the simplified and full Maxwell equation solutions has reduced to $10^{-2}$. This indicates that any effect of covariant terms is local and would not accumulate if we propagated the wave through multiple overdensities. Note lastly that at this time stamp, the constraint violation has also dropped by an order of magnitude so that the deviation between the simplified and full Maxwell equation solutions is still an order of magnitude above the constraint violation.
\newline \newline
\begin{figure*}
\includegraphics[width=13cm]{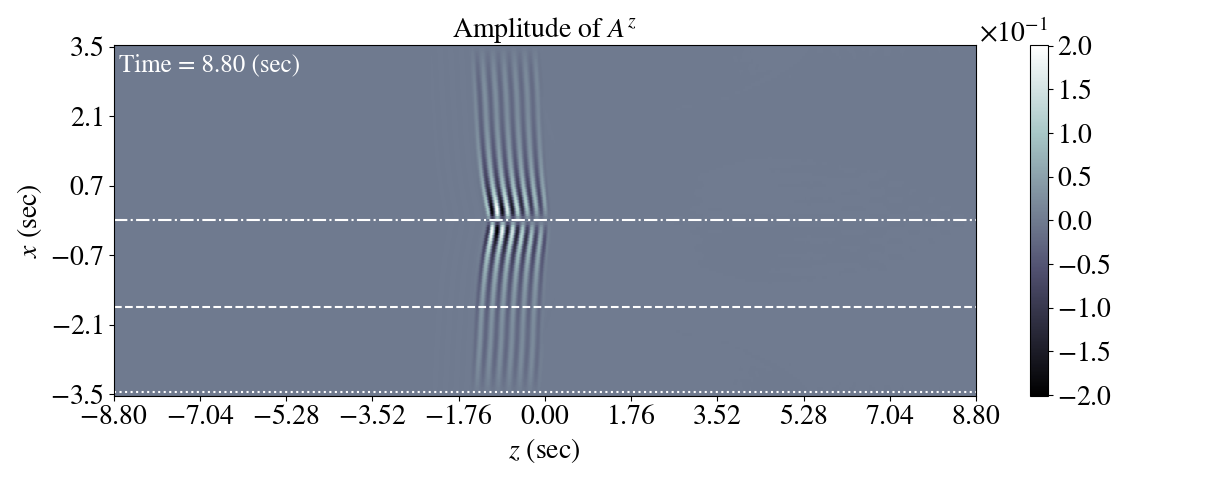}
\includegraphics[width=13cm]{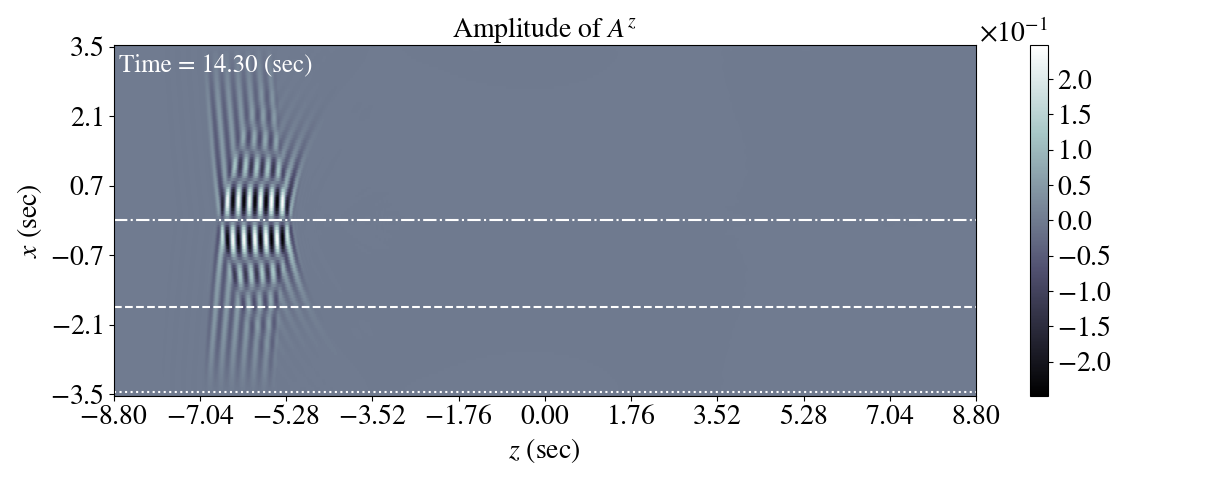}
\caption{Evolution of $A^z$ for the simulation with $|\vec{k}| 1.5M_\textrm{gi}$. We do not track the amplitude of wave-fronts in these plots. The center of the dark matter particle is at the origin.}
\label{evol_kgM_Az}
\end{figure*}
Before moving on to discuss the second simulation, we provide a few points regarding the evolution of $A^z$ and $\phi$. Both of these components are initialized as zero but develop non-vanishing amplitudes as the wavefront approaches the origin. The maximum amplitude reached is of order $10^{-1}$. This magnitude persists even after the wavefront has passed through the dark matter particle. We show this for $A^z$ in figure \ref{evol_kgM_Az} and note that the result for $\phi$ is similar (and thus not shown). Since this amplitude is (mostly) larger than the constraint violation, we expect this may be a genuine effect, i.e. we expect the wavefront may genuinely develop a longitudinal component. This is supported by several other considerations. Firstly, we note that $A^z$ and $\phi$ both remain negligible in the simplified Maxwell equation case even though this case has stronger constraint violation since we are dropping terms. Furthermore, we can understand the appearance of the non-vanishing $A^z$ and $\phi$ by looking at Maxwell's equations (\eqref{Az}, \eqref{phi}). In these we see that the evolution equations for $A^z$ and $\phi$ contain covariant terms proportional to $A^x$ which lead to an evolution of $A^z$ and $\phi$ away from zero. However, we also note that the authors of \cite{keller2019electrodynamics}, mention that when the Lorenz gauge condition (the constraint violation) is non-zero, it could result in longitudinal polarization. To test if the non-vanishing $A^z$ and $\phi$ is indeed an artifact we would need to increase the precision of our simulations by moving beyond the linear approximation.
\newline \newline
Lastly, we remind the reader that we have run a modified version of the simulation where the spline was not introduced in $\alpha$. This simulation leads to the same overall results as those presented here but with stronger constraint violation initially.

\subsection{Simulation with $|\vec{k}| = 0.1M_\textrm{gi}$}
\begin{figure*}
\includegraphics[width=16cm]{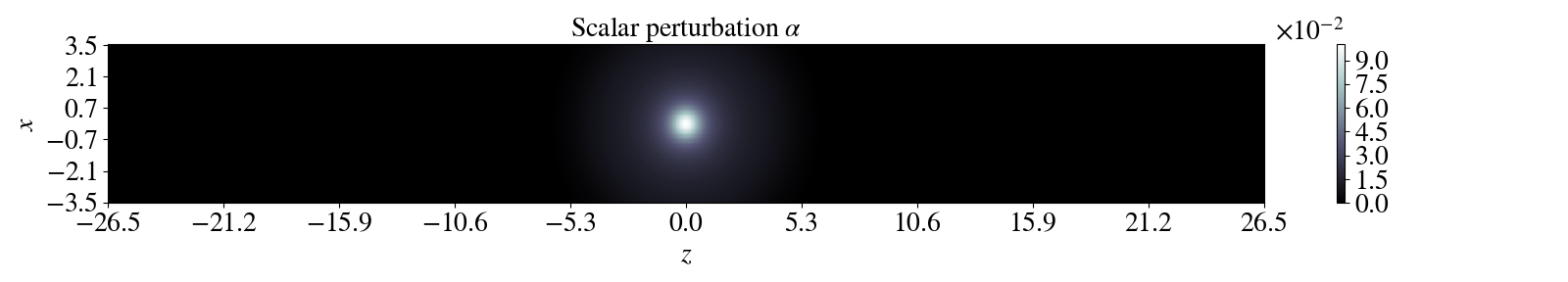}
\includegraphics[width=16cm]{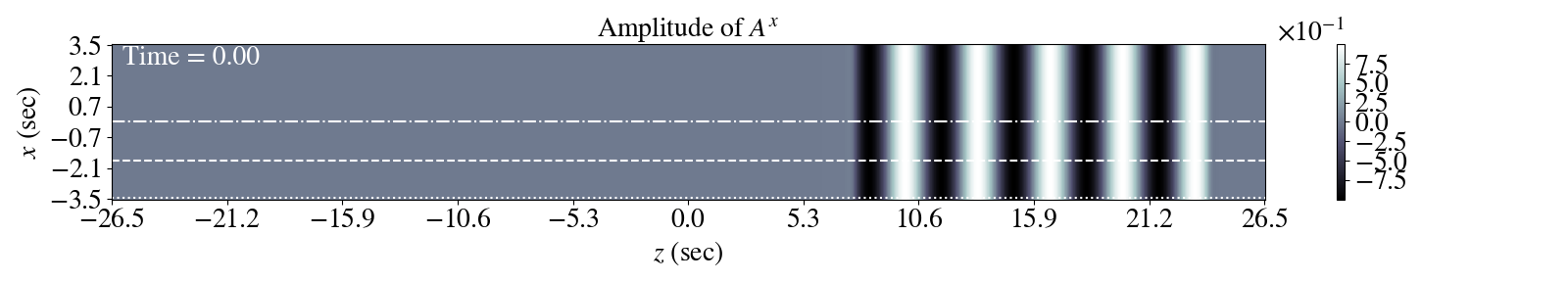}
\includegraphics[width=16cm]{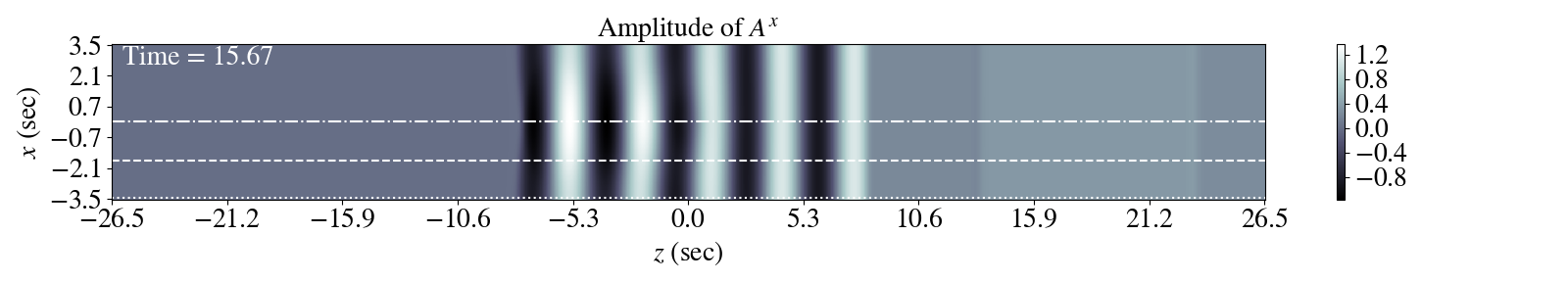}
\includegraphics[width=16cm]{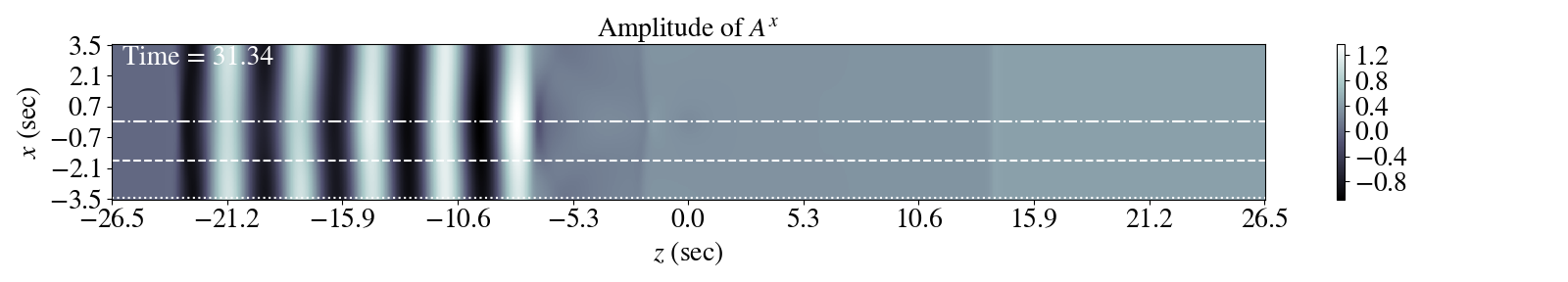}
\caption{The top figure shows the scalar perturbation $\alpha$ in the $xz-$ plane. The bottom three plots show the evolution of $A^x$ for the simulation with $|\vec{k}| = 0.1M_\textrm{gi}$. The plots include three horizontal lines parallel to the z-axis and at different distances in the x-axis, indicating where we track the maximum amplitude of wave-fronts as shown in figure \ref{amp,kgM}.}
\label{evol_klM}
\end{figure*}

In this subsection we discuss the results from propagating a wavefront with $|\vec{k}| < M_\textrm{gi}$ through the simulation box.
\newline \newline
Figure \ref{evol_klM} shows the wavefront at three instances: At initial time, when the wavefront is centered at the origin and when the wavefront has passed entirely through the central overdensity. In this simulation we do not see the same interference pattern that we saw in the earlier simulation. Despite the lack of interference pattern, we do see a minor increase in the amplitude of the wave as it propagates through the inhomogeneous region. In addition, the figure shows that part of the wavefront is reflected off the dark matter particle which may be related to the findings of \cite{koksbang2022effect} where it was found (within certain assumptions) that a wavefront cannot propagate through a steep overdensity if $|\vec{k}| < M_\textrm{gi}$. However, we also observe reflections of the wave when evolving the simplified Maxwell equations (results not shown). We will comment further on this below when comparing to earlier work.
Although the interference pattern in this simulation is much more subtle than in the former simulation, we do note that a close inspection reveals that the peak amplitudes of each full wavefront are not constant and that the wave is thus no longer a simple plane wave. This can both be due to the interference induced by the Shapiro time delay, and the reflection of part of the wave.
\newline

\begin{figure}
    \centering
    \includegraphics[width=1\linewidth]{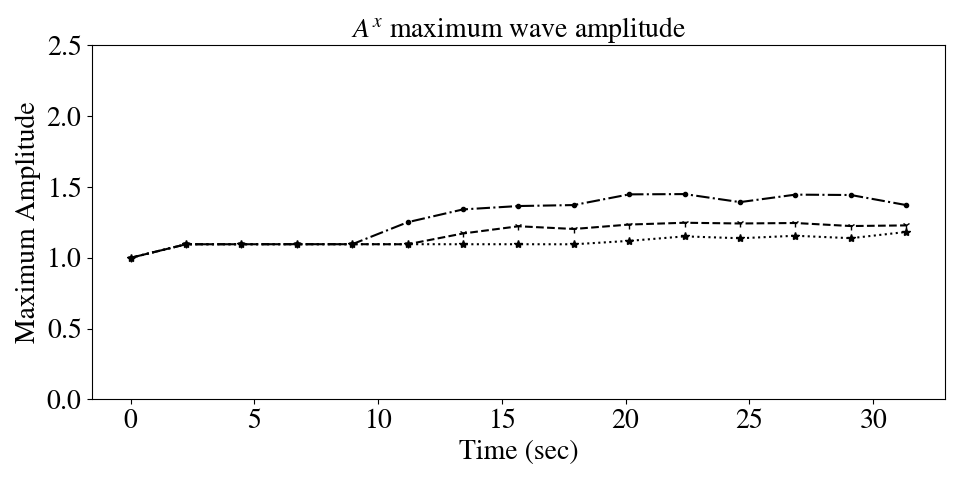}
    \caption{Maximum amplitudes of the wave as it propagates through the dark matter particle in the simulation with $| \Vec{k} | = 0.1 M_\textrm{gi}$. The different line types corresponds to the horizontal lines in figure \ref{evol_klM} where we track the maximum amplitude of wave-fronts.}
    \label{amp,klM}
\end{figure}
\begin{figure*}
\includegraphics[width=15.96cm]{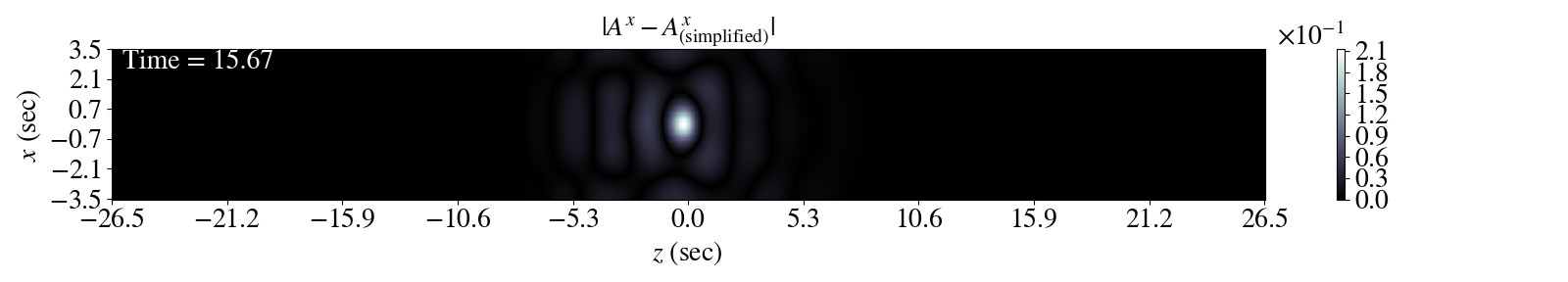}
\includegraphics[width=15.96cm]{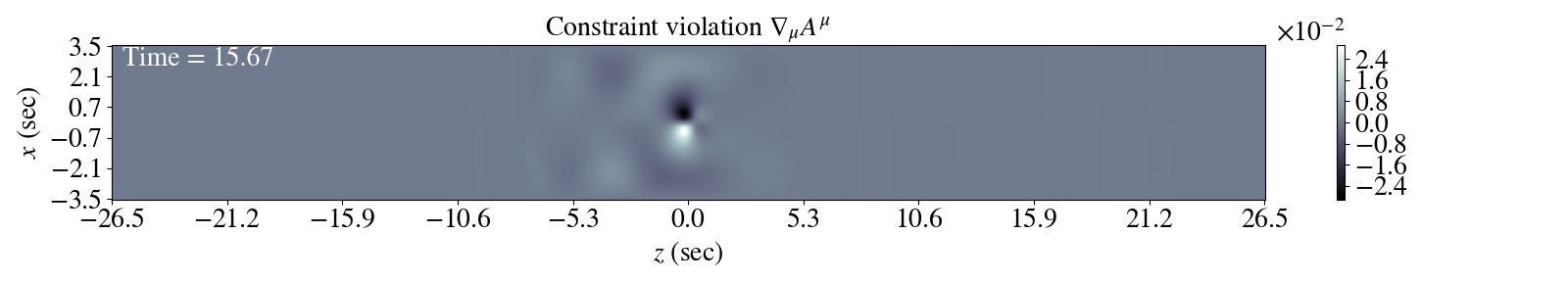}
\includegraphics[width=15.96cm]{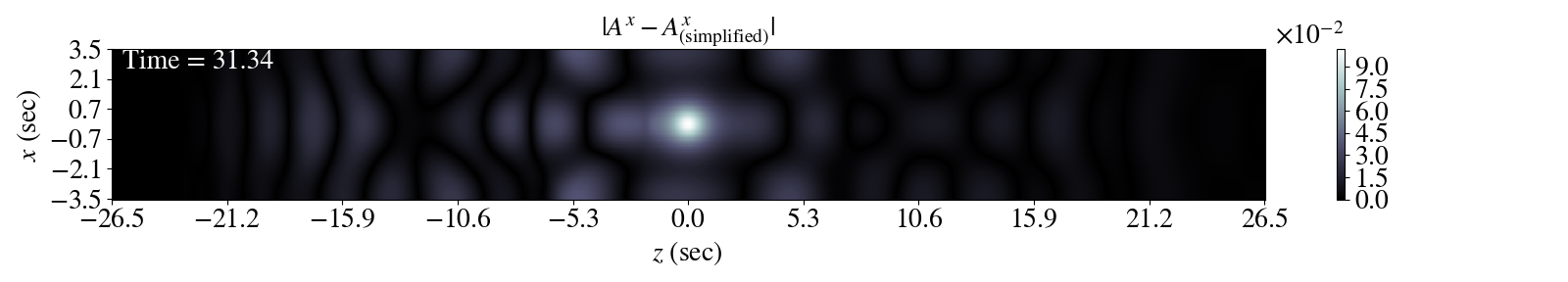}
\includegraphics[width=15.96cm]{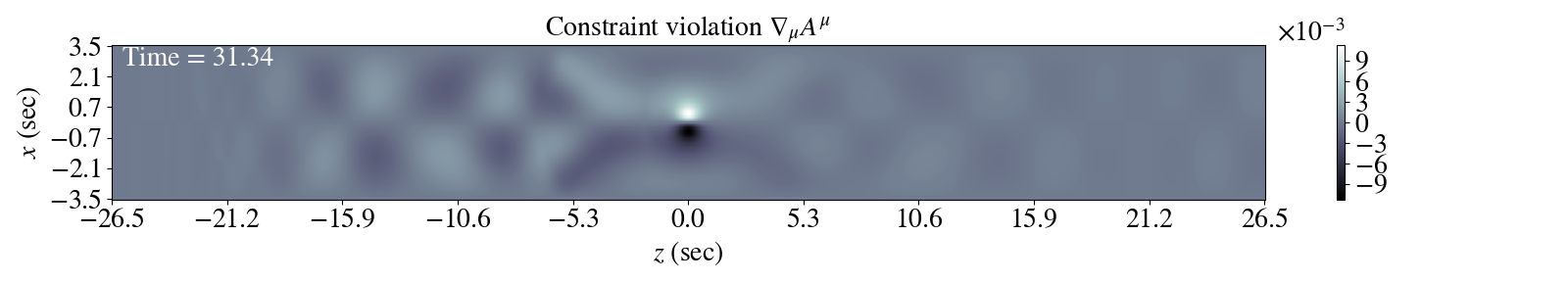}
\caption{The absolute difference between $A^x$ (eq. \eqref{Ax}) and the simplified Maxwell equation $A^x$, and its corresponding constraint violation at two time instances for simulation with $|\Vec{k}| = 0.1M_\textrm{gi}$. The center of the dark matter particle is at the origin. Constraint violation plots are for simulations with full Maxwell's equation.}
\label{nc_klM}
\end{figure*}
\begin{figure*}
\includegraphics[width=15.96cm]{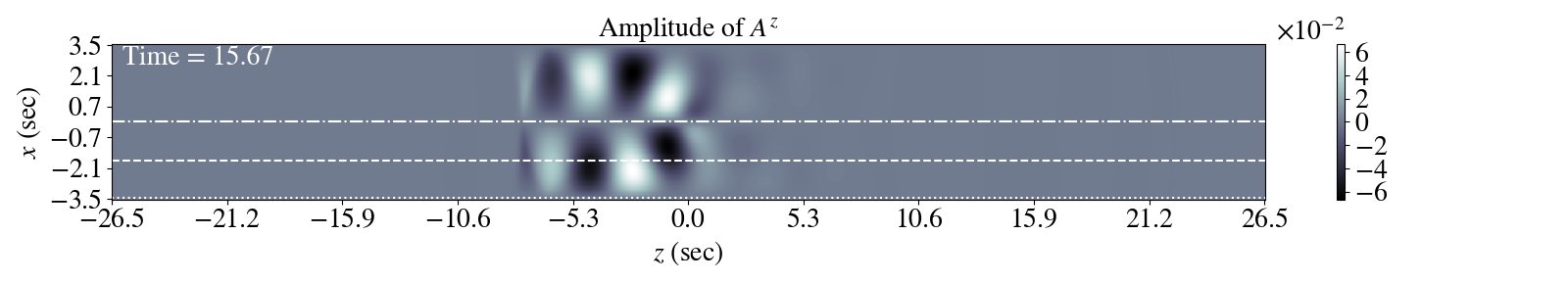}
\includegraphics[width=15.96cm]{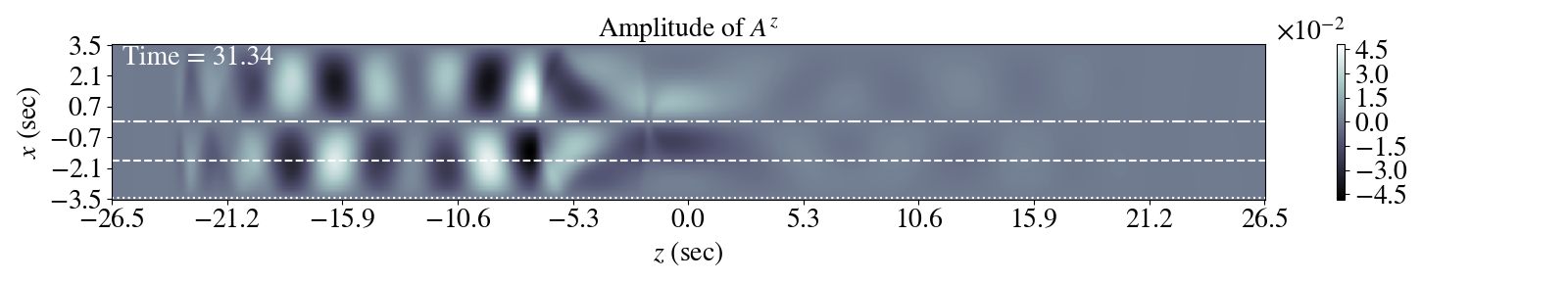}
\caption{Evolution of $A^z$ (despite having zero initial amplitude) for simulation with $|\Vec{k}| =0.1M_\textrm{gi}$. We do not track the amplitude of wave-fronts in these plots. The center of the dark matter particle is at the origin. Spatial distances and time are in units of seconds.}
\label{evol_klM_Az}
\end{figure*}

Similarly to the simulation with $| \vec{k} | = 1.5M_\textrm{gi}$, figure \ref{amp,klM} shows the amplitude of the wavefront at three different distances corresponding to the three lines shown in figure \ref{amp,klM}. The figure shows that the amplitude does not change as dramatically here as it did in the earlier simulation.
\newline \newline
As with the other simulation, we find that the absolute difference between $A^x$ in the simplified and full Maxwell equation case deviates by a few orders of magnitude, but with a very different pattern than what we found in the previous simulation. This is shown in figure \ref{nc_klM}. When looking at the absolute difference between $A^x$ in the simplified versus full Maxwell equation case, we find the difference to be of the order of $10^{-1}$ when the wave is at the origin and order of $10^{-2}$ after the wave has left the dark matter particle. We also note that the constraint violation remains in the order of $10^{-2}$ for both the time instances, when the wave is at the origin and when it has left the dark matter particle. Overall, this simulation is hard to interpret because of the reflection of the wave which leads to interference and in general to the wave no longer being plane. 
\newline \newline
Lastly, we again note that both $A^z$ and $\phi$ were initialized as zero but developed non-vanishing amplitudes as the wavefront approached the y-axis. Their maximum amplitude is of the order $10^{-2}$. This magnitude persists even after the wavefront has passed entirely through the dark matter particle. However, in this case the constraint violation is of the same order of magnitude as $A^z$ and $\phi$. Thus, any genuine development of $A^z$ and $\phi$ is in this case below the precision of the linear approximation. We show $A^z$ in figure \ref{evol_klM_Az} along with the constraint violation and note that the result for $\phi$ is similar (and thus not shown).
\subsection{Comparison with earlier work}
In this section, we compare our results with earlier work considering electromagnetic waves in curved spacetime. 
\newline\newline
In \cite{PhysRevD.3.1708}, the authors solve Maxwell's equations analytically in a static weak gravitation field. Focusing on high-frequency waves in a spacetime with a weak spherically symmetric gravitational field, the authors find that the gravitational field increases the amplitude of the light near the mass. This is in agreement with our findings, where we for the shorter wavelength also find an increase in the wave amplitude by approximately a factor of four in the light propagating through the center of the overdensity compared to the light propagating far from it. The authors of \cite{PhysRevD.3.1708} also find that there is no change in the transverse polarization but some change in the longitudinal component when light propagates through a spherically gravitational field. Our results again agree with this. Specifically, $A^y$ remains negligible throughout our simulations, while we do see minor developments in $A^z$.
\newline\indent
The results of \cite{Turyshev_2017} are very similar. The authors study high-frequency light propagating around the Sun modeled as a spherically symmetric overdensity in the post-Newtonian approximation. The authors find that the Sun's gravitational field amplified the amplitude of the light but that the polarization does not change. This is in agreement with our numerical findings.
\newline\indent
In \cite{koksbang2022effect}, the authors introduced a beyond-geometric optics approximation where the Ricci-term was of the same order as the otherwise dominant term in Maxwell's field equation. The beyond-geometric optics scheme was then analyzed in a setting similar to ours, where light propagates through a classical matter distribution mimicking a dark matter particle. In this scheme, light rays were still found to be transverse, but not in general null while traveling through the dark matter particle. Most intriguingly, the authors also found that light cannot enter the region of the dark matter particle if $|\vec{k}|< M_\textrm{gi}$. These new results were all due to the Ricci term. In our simulations, the Ricci term is only around 1 percent of the other terms and thus we do not expect the see the results discussed in \cite{koksbang2022effect}, and indeed, we only see minor effects of the curvature term while the main deviations we see from geometric optics are due to wave optics. However, the reflection we see in the $|\vec{k}|<M_\textrm{gi}$ case  when the wave reaches the overdensity may be the small-curvature limit of the prediction in \cite{koksbang2022effect} that the wave cannot propagate through the overdensity if the curvature is strong enough.
\newline\indent
In \cite{Asenjo_2017}, the authors find exact null-geodesic solutions to Maxwell's equations for static spherically symmetric spacetimes, although in a very special coordinate system. At least naively, this seems contrary to our results since we find an index of refraction different than 1 when light propagates through the overdensity. However, a direct comparison between our results and those of \cite{Asenjo_2017} is difficult due to the special coordinate choice of \cite{Asenjo_2017}.
\newline\newline
We lastly note that from a formal point of view, our simulations are similar to some studies on gravitational waves. Specifically, we note that \cite{he2021gwsim} presents a code for propagating gravitational waves through a scalar gravitational potential using a simplified Maxwell's equation in the perturbative limit. In agreement with our findings, the results of \cite{he2021gwsim} show that the waves are retarded by the scalar perturbation, leading to an interference pattern and increased amplitude similar to what we find in our first simulation.


\section{Summary, discussion and conclusions} \label{sec:summary and conclusions}
We solved Maxwell's equations to linear order in the perturbation 
for a light wave propagating through a scalar potential mimicking the spacetime in and around a dark matter particle. We found that the scalar potential induces an interference pattern due to the Shapiro time delay, i.e. the interference pattern appears because the scalar gravitational potential locally slows down the speed of light. For one of our simulations, the interference pattern includes regions where the wave amplitude is consistently approximately four times larger than the original amplitude. Since the effect is persistent we would expect an accumulation if a wavefront propagates through several dark matter particles, but note that we would need to expand our current numerical setup to include more structures if we were to determine to what extent this accumulation actually appears. For the other simulation, the amplification is much more modest and since we obtain different amplifications for different frequencies of light for a given curvature, we find that the amplification factor could depend on the frequency of light. Further, in this simulation the overall interference pattern is more complicated. The complication arises because part of the wave front is reflected on the scalar potential. This reflection may be a small-curvature version of the result in \cite{koksbang2022effect} where it was found that light rays with sufficiently long wavelengths cannot propagate into regions corresponding to dark matter particles described by classical mass distributions.
\newline\indent
We aim at adapting our numerical code to go beyond the linear regime and indeed to work with general spacetimes. It would be particularly interesting to consider an exterior Schwarzschild spacetime to e.g. complement analytical studies such as \cite{murk2024gravityinducedbirefringencesphericallysymmetric}. Future work could also expand the current code to Kerr and Godel spacetimes. Rotating spacetimes structures are particularly interesting since real black holes rotate and at the same time, the analytical (approximate) considerations in \cite{Asenjo_2017} find that the rotations lead to light rays being neither null nor geodesic  \cite{Frolov_2012,Frolov_2020}. It would be very interesting to see the corresponding numerical results without introducing approximations. This may prove important for validating the accuracy of studies that aim at testing quantum gravity with black hole images such as the line of work addressed in \cite{carballorubio2024disentanglingphotonringsgeneral}. Another interesting situation that could benefit from numerical investigation is the suggestion in 
\cite{asenjo2023abnormallightpropagationunderdetermination} that astrophysical observations might not be (entirely) based on parallel plane waves. The authors there demonstrate other options which it would be interesting to study in numerical setups. We also note that our numerical simulations to a limited extent encourages the study of beyond-plane wave observations since our simulations show that the scalar potential induces an intricate interference pattern that does not in general correspond to a locally parallel plane wave.
\newline\indent
Overall, we believe that astrophysical observations have reached a precision where beyond-geometric optics effects may be important. Numerical studies of solutions to Maxwell's equations is one tool for studying these effects in controlled environments and where it is feasible to move beyond the simplifying approximations and assumptions necessary for analytical studies.


\begin{acknowledgments}
APS and SMK are funded by VILLUM FONDEN, grant VIL53032 (PI: SMK). Research at Perimeter is supported by the Government of Canada through the Department of Innovation, Science, and Economic Development, and by the Province of Ontario through the Ministry of Colleges and Universities. The computational work was performed using the DeiC Large Memory HPC System (Hippo) managed by the eScience Center at the University of Southern Denmark.
\newline\indent
The presented work is based on early considerations between SMK and Syksy Rasanen and the authors thank Syksy Rasanen for correspondence. APS and SMK also thank Shouryya Ray for pointing us towards Cadabra\footnote{https://cadabra.science/}, an opensource computer algebra software package \cite{Peeters2018,peeters2018introducingcadabrasymboliccomputer,Peeters_2007} which was used to verify analytical derivations. APS thanks Lucas Timotheo Sanches for help with setting up the Einstein Toolkit in the Hippo HPC system.
\newline\newline
{\bf Author contribution statement}: APS and SMK performed the analytical computations. APS performed the numerical work under the guidance of SMK and ES. Analysis of the numerical work was led by APS but all authors contributed. Conclusions were drawn by APS and SMK who jointly wrote the manuscript with comments from ES.
\end{acknowledgments}

\appendix
\begin{figure}
    \centering
    \includegraphics[width=1\linewidth]{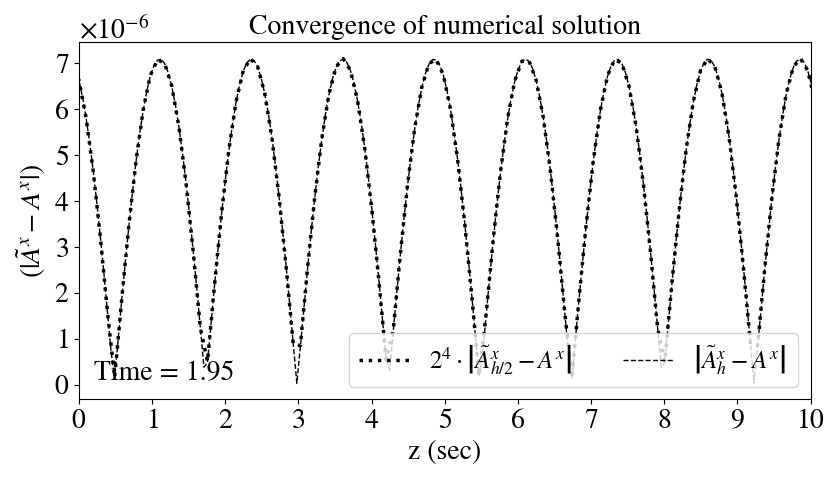}
    \includegraphics[width=1\linewidth]{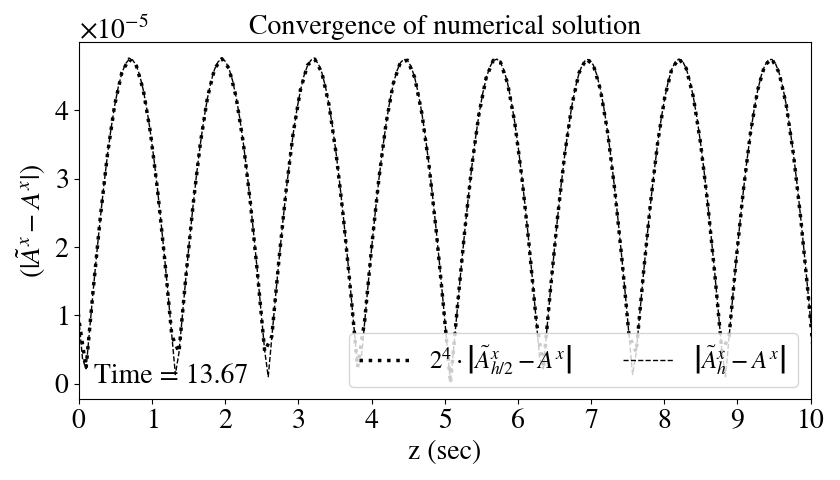}
    \includegraphics[width=1\linewidth]{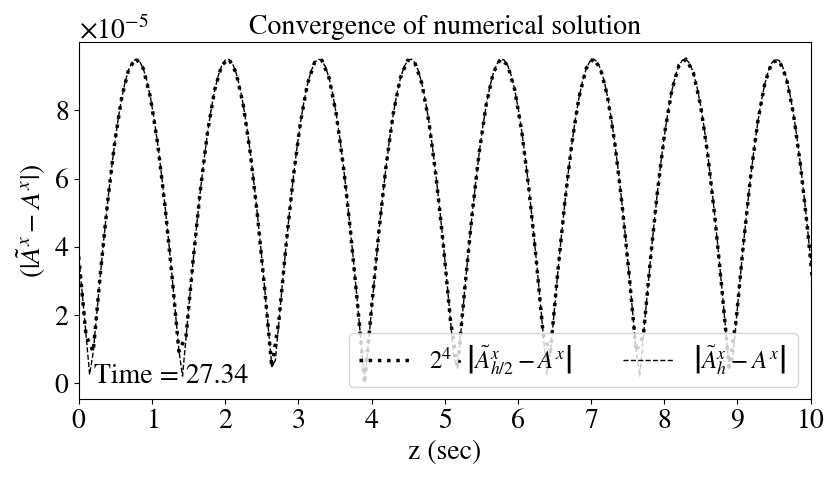}
    \caption{Convergence plots at three different time instances for Maxwell equations in flat spacetime for resolutions - ($64 \times 64 \times 1024$) and ($128 \times 128 \times 2048$) denoted by $h$ and $h/2$ respectively. The data was sampled along a line from $(0,0,0)$ to $(0,0,10)$. For fourth order convergence, we expect the high resolution error to be $16$ times lower than the low resolution error; hence, we multiply the high resolution error by a factor of $2^4$ to show numerical convergence.}
    \label{convergence_plot}
\end{figure}
\section{Convergence of numerical solutions}\label{app:convergence}
To show that the numerical simulations used for this work converge, we here present the result of the comparison between numerical and analytical solutions of Maxwell's equation in flat spacetime. In this section, the variables $A^x$ and $\tilde{A^x}$ denote analytic and numerical solutions, respectively. We perform two simulations with grid resolutions $64 \times 64 \times 1024$ and $128 \times 128 \times 2048$, denoted by $h$ and $h/2$ respectively. We initialize a plane wave propagating along the z-axis, i.e. initial conditions are set according to

\begin{equation}\label{eq:planewave_flat}
    \begin{split}
        A^x |_{(t=0)} & = \cos( 2 \pi k_zz )\\
        A^y|_{(t=0)} & = 0 \\ 
        A^z|_{(t=0)} & = 0 \\
        \phi|_{(t=0)} & = 0.
    \end{split}
\end{equation}

We use periodic boundary conditions in all three spatial dimensions. We set the domain length as $L = 50$ sec and $k_z = 0.4 \text{ sec}^{-1}$. With eqn. \eqref{eq:planewave_flat} and periodic boundary conditions, e.g. $A^x(x + L) = A^x(x)$, we find the analytic solution,

\begin{equation}
    A^x = \cos\left(\frac{2 \pi k}{L} z - \frac{2 \pi k}{L} t\right).
\end{equation}

We then compute the absolute difference between the numerical and analytical solution, $|\tilde{A^x} - A^x|$, for both the resolutions at three time instances. We verify the order of convergence, $p$, which should be $p=4$ in our case. Therefore,

\begin{equation}\label{conv_eqn}
\begin{split}
    \left| \frac{\tilde{A^x}_h - A^x}{\tilde{A^x}_{h/2} - A^x} \right|  &= 2^p + O(h).\\ 
    \text{or, } |\tilde{A^x}_h - A^x| & \simeq 2^4  |\tilde{A^x}_{h/2} - A^x| .
\end{split}
\end{equation}

The figure \ref{convergence_plot} compares the error between two different resolutions that agree with eqn. \eqref{conv_eqn} and shows that solutions converge by order of $4$.

\bibliography{ref}


\end{document}